\documentclass[fleqn,10pt]{wlscirep}
\title{Toward Fast Neural Computing using All-Photonic Phase Change Spiking Neurons}

\author[1,*]{Indranil Chakraborty}
\author[1]{Gobinda Saha}
\author[1]{Abhronil Sengupta}
\author[1]{Kaushik Roy}
\affil[1]{Purdue University, School of Electrical and Computer Engineering, West Lafayette, IN-47907, USA}

\affil[*]{ichakra@purdue.edu}
\usepackage{soul}
\begin{abstract}
The rapid growth of brain-inspired computing coupled with the inefficiencies in the CMOS implementations of neuromrphic systems has led to intense exploration of efficient hardware implementations of the functional units of the brain, namely, neurons and synapses. However, efforts have largely been invested in implementations in the electrical domain with potential limitations of switching speed, packing density of large integrated systems and interconnect losses. As an alternative, neuromorphic engineering in the photonic domain has recently gained attention. In this work, we propose a purely photonic operation of an Integrate-and-Fire Spiking neuron, based on the phase change dynamics of Ge$_2$Sb$_2$Te$_5$ (GST) embedded on top of a microring resonator, which alleviates the energy constraints of PCMs in electrical domain. We also show that such a neuron can be potentially integrated with on-chip synapses into an all-Photonic Spiking Neural network inferencing framework which promises to be ultrafast and can potentially offer a large operating bandwidth.
\end{abstract}
\begin{document}

\flushbottom
\maketitle
\thispagestyle{empty}

% \noindent Please note: Abbreviations should be introduced at the first mention in the main text – no abbreviations lists. Suggested structure of main text (not enforced) is provided below.

\section*{Introduction}

The recent advances in the field of neuromorphic computing largely rest on our understanding  
of the human brain as researchers strive to comprehend the intricacies of its complex functionalities and emulate its unparalleled energy efficiency. Despite the obvious elusivenss of the brain, neuroscientific experiments have unravelled various underlying mechanisms behind our behavorial patterns. To that effect, various studies have been performed exploring phenomena concerning the basic functional units, namely neurons and synapses, that knit the neural network in the human brain. The need to incorporate these neuroscientific findings in computing models and consequently in building bio-plausible hardware has led to extensive investigations in recent years. \par
Most of the available computing models that encode the information processing in a neural network are based on mathematical optimization techniques. More recently, with growing evidence of spike-based processing in the biological neural network, its event-driven nature has led researchers to explore bio-plausible hardware implementations in an effort to achieve higher energy efficiency. Spiking neural networks (SNN) comprise the third generation of neural networks and the basic principle relies on how the membrane potential of a spiking neuron rises and eventually cause the neuron to spike under the action of incident spikes. Hardware implementations of various spiking neuron models such as Hodgkin-Huxley\cite{Hodgkin_1952} and Leaky-Integrate-Fire (LIF)\cite{Stein_1967} on CMOS platforms not only fail to match the energy efficiency of the human brain but is also area-inefficient.\par 
To address these shortcomings, novel material systems and technologies \cite{Tuma_2016, sengupta2017encoding} have been proposed to mimic the behavior of a spiking neuron thus providing direct mapping between a single device behaving as a functional neural element. However, each technology suffers from different drawbacks, such as energy-efficiency, speed, cross-talk, fabrication difficulties, etc. Phase change materials (PCM), in particular, have been demonstrated \cite{rajendran2013specifications} to have significant energy restrictions due to their high `write' times in the electrical domain. It has been shown that either the exciting current or `write' pulse duration has to be reduced by 10$\times$ for PCM to perform better than CMOS. However, recently PCMs, e.g. GST, have been demonstrated\cite{Stegmaier_2016} to achieve sub-ns `write' speeds when excited by photonic laser pulses. Due to highly contrasting optical and electrical properties in their amorphous (a-GST) and crystalline (c-GST) states, PCMs have thus offered avenues to implement all-photonic memories\cite{pernice2012photonic, Rios_2015}, switches\cite{Stegmaier_2016} and have been even used for mixed-mode electro-optical operations \cite{Rodriguez_Hernandez_2017}. The promise of fast information processing with PCMs in the photonic domain has thus encouraged the possibilities of PCMs as a viable material for photonic neuromorphic systems. Recently, device\cite{cheng2017chip} based on GST deposited on waveguides was proposed to emulate the synaptic weight update mechanism in synapses in SNN framework. Previous works on such based spike-based neuromorphic processing in the photonic domain have been dependent on electro-optic conversions\cite{tait2014broadcast,fok2012asynchronous} where lasers have been used to emulate the behavior of spiking neurons. In this paper, we propose an all-photonic operation of an Integrate-and-Fire spiking neuron. We show that the proposed neuron mimics the behavior of the biological neuron and can be seamlessly integrated in an all-photonic SNN framework. Other works in the photonic neuromorphic domain includes applications such as deep neural networks \cite{Shen_2017} and recurrent neural networks\cite{Tait_2017} which are complementary to the processing framework discussed in this paper.\par
\section*{GST embedded Ring Resonator as a Integrate-Fire Neuron}

\begin{figure*}[t]
		\centering
		\includegraphics[width=0.8\textwidth,keepaspectratio]{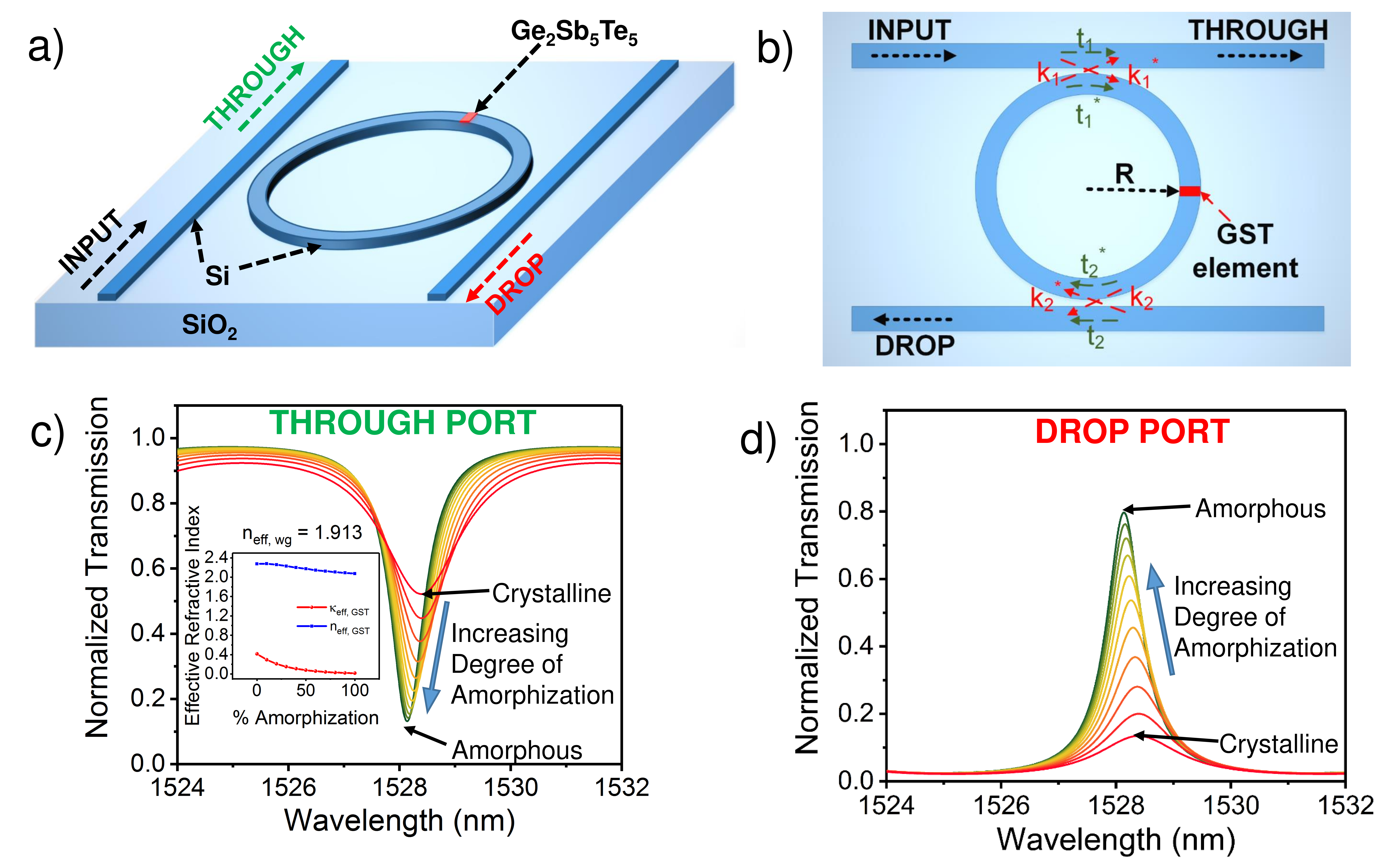}
		\vspace{-2 mm}
		\caption{(a) A perspective view of an add-drop microring resonator with a small patch of GST on top showing its ports and materials. (b) A two-dimensional top view of the ring resonator illustrating the input, output, coupling and transmission parameters. Theoretically calculated transmission at various wavelengths for different degrees of amorphization of GST ranging from 0\% (crystalline) to 100\% (amorphous) showing that the transmission at the (c) `THROUGH' ((d) `DROP') port decreases (increases) with increasing degree of amorphization. }
		\vspace{-4mm}
		\label{fig:theory}
	\end{figure*}

The basic working principle of a ring resonator is necessary to be illustrated at first. A ring resonator is a structure with two rectangular waveguides and a ring waveguide (as shown in Fig. \ref{fig:theory} (a)). Wave entering through the `INPUT' port gets partially coupled to the ring waveguide and interferes constructively inside the ring when the following condition, called resonant condition, is met:
\begin{equation}
2\pi Rn_{eff,wg} = m\lambda_m
\end{equation}
Eq.1 provides the resonant condition (at wavelengths $\lambda_m$) for the ring resonator of radius $R$ where the effective refractive index of the waveguide-substrate material system is $n_{eff,wg}$. By controlling the coupling and attenuation parameters, $t_1, t_2$ and $k_1, k_2$, as shown in Fig. \ref{fig:theory} (b), light can be conditionally guided through the `THROUGH' and `DROP' ports. \par
Introducing a GST element (shown in red in Fig. \ref{fig:theory}(a)) on top of the ring waveguide in the ring resonator described above allows us to control light propagation through the ports by merely changing the state of the GST. Light passing through the waveguide get evanescently coupled to the GST element and gets differentially absorbed by the GST in its low-loss amorphous state and high-absortion crystalline state \cite{pernice2012photonic}. The difference in attenuation arises due to the contrasting imaginary refractive index ($\kappa_{GST}$) of GST in its two states. Theoretically, the transmission of the `THROUGH' and `DROP' ports can be expressed as:

\begin{equation}
T_t = \frac{t_2^2\alpha^2-2t_1t_2\alpha cos(\theta)+t_1^2}{1-2t_1t_2\alpha cos(\theta)+(t_1t_2\alpha)^2}\\
T_d = \frac{(1-t_1^2)(1-t_2^2)\alpha}{1-2t_1t_2\alpha cos(\theta)+(t_1t_2\alpha)^2}
\end{equation}
where $\alpha$ is the attenuation factor, $\theta$ is the phase factor, $t_1$ and $t_2$ are coupling parameters. $\alpha$ and $\theta$ can be expressed as:
\begin{flalign}
\alpha &= exp(-\frac{2\pi}{\lambda}[\kappa_{eff,wg}(2\pi R-L_{GST})+\kappa_{eff,GST}L_{GST}])\approx exp(-\frac{2\pi}{\lambda}\kappa_{eff,GST}L_{GST})&&\\
\theta &= \frac{2\pi}{\lambda}[n_{eff,wg}(2\pi R-L_{GST})+n_{eff,GST}L_{GST}].
\end{flalign}
Here $\kappa_{eff,GST}$ ($\kappa_{eff,wg}$) and $n_{eff,GST} $($n_{eff,wg}$) are effective imaginary and real parts of the refractive index of the waveguide material with (without) GST. $R$ is the radius of the ring waveguide and $L_{GST}$ is the length of the GST element. 
The refractive indices of partially crystallized GST are estimated from effective permitivities approximated by an effective-medium theory\cite{Chen_2015,Voshchinnikov_2007}:
\begin{equation}
\frac{\epsilon_{eff}(p)-1}{\epsilon_{eff}(p)+2}=p\times\frac{\epsilon_{c}-1}{\epsilon_{c}+2}+(1-p)\times\frac{\epsilon_{a}-1}{\epsilon_{a}+2}
\end{equation}
where $\epsilon_{c}$ and $\epsilon_{a}$ are the permittivities in the crystalline and amorphous states respectively calculated from the refractive indices of GST\cite{pernice2012photonic} by $\sqrt{\epsilon(\lambda)}=n+i\kappa$. $p$ is the degree of crystallization. 
The effective refractive indices of the Si waveguide- SiO\textsubscript{2} substrate system with and without GST was calculated using COMSOL Multiphysics simulations, shown in the inset of Fig. \ref{fig:theory} (c). 
\begin{figure*}[t]
		\centering
		\includegraphics[width=1\textwidth,keepaspectratio]{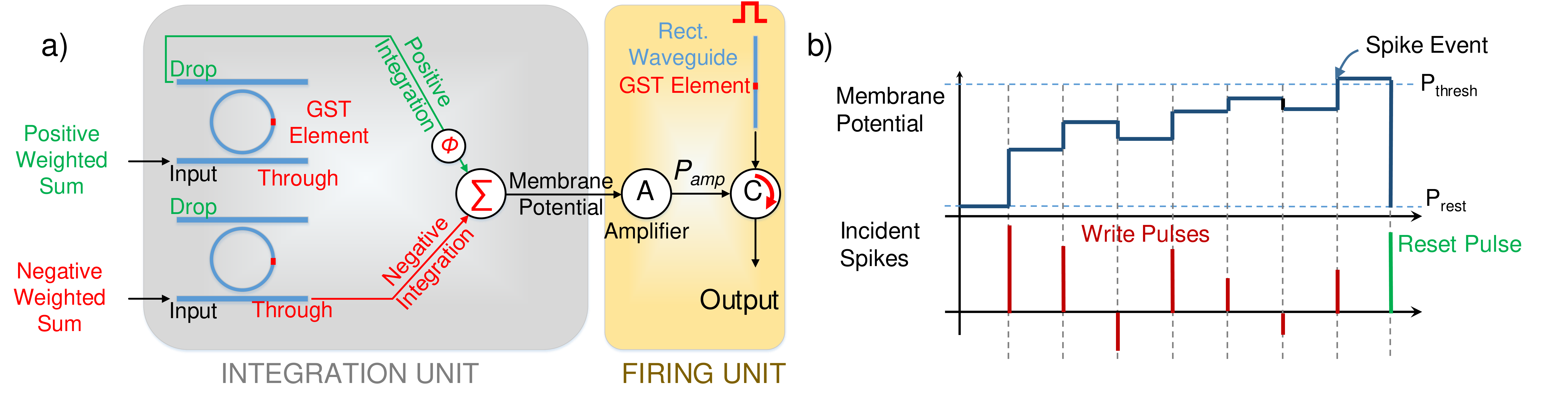}
		\vspace{-2 mm}
		\caption{(a) Schematic of a bipolar integrate and fire neuron based on GST-Embedded Ring resonator devices showing the integration and firing unit. (b) Timing diagram showing the integration of membrane potential for various incident pulses demonstrating the operation of the proposed neuron.}
		\vspace{-4mm}
		\label{fig:neuron}
	\end{figure*}
These equations depict the theoretical backdrop of a ring resonator system with GST. As the GST element crystallizes (amorphizes), $\kappa_{eff,GST}$ and hence its absorption increases (decreases) and as a result the transmission at the `THROUGH' (`DROP') port increases. Fig. \ref{fig:theory} (c) and \ref{fig:theory} (d) shows that the theoretically calculated transmission at the THROUGH and `DROP' ports in a ring resonator increases with $p$. We propose an integrate-fire spiking neuron leveraging these characteristics of the GST-ring resonator system. \par
Information processing in neural networks usually involve multiplication of inputs with the significance metric of the synapses, namely `weight' and feeding the corresponding output to a neuron. For most neural network applications, weights can assume negative values. It is thus necessary to realize a bipolar neuron which can receive inputs of both polarity for all practical purposes. Let us now consider a GST embedded ring resonator described above. The GST initially is in crystalline state, denoting the highest (lowest) transmission level through `THROUGH' (`DROP') port. During the `write' phase, an off-resonance pulse is input which writes into the GST element, thereby reducing (increasing) its degree of crystallization $p$ (amorphization ($1-p$)). During the `read' phase, as $p$ reduces, `THROUGH' port transmission $T_t$ decreases and `DROP' port transmission ($T_d$) increases. Thus, with incoming pulses, the transmission through the `DROP' and `THROUGH' ports get positively and negatively integrated respectively. We combine these properties of the device to propose a bipolar integrate and fire neuron. The integration unit of the neuron body consists of two ring resonators as shown in Fig. \ref{fig:neuron} (a) and pulses of amplitudes proportional to the positive ($O_j^+$) and negative ($O_j^-$) weighted sums, received from the synapses, are fed to the positive and negative ring resonators respectively. The details of entire network framework is discussed later. Note, the resultant amplitude of the incident pulse to the neuron is the difference of the positive and negative inputs fed to the two devices: $O_j = O_j^+ - O_j^-$. Thus, the two ring resonators integrate in opposite direction to emulate the resultant integration which should ideally be proportional to $O_j$. 
The output from the `DROP' and `THROUGH' ports of the positive and negative devices respectively are passed to an interferometer. We place a phase modulator ($\phi$) in the path of the positive ring resonator and the interferometer to tune the output of the interferometer to produce the sum of the two incoming pulses. As the two ports integrate in the opposite direction, the output of the interferometer is the resultant integration based on both the positive and negative inputs to the neuron body and can be treated as membrane potential of the integrate and fire neuron. Thus at every time-step, the membrane potential of the $j^{th}$ neuron can be represented by: 
\begin{equation}
V_j[t] = V_j[t-1] + O_j[t]
\end{equation}
Fig. \ref{fig:neuron} (b) shows the operation of the proposed neuron such that the membrane potential integration is proportional to the amplitude of the resultant incident spike to the neuron. Once the GST reaches full amorphization, the membrane potential crosses its threshold ($P_{thresh}$). The `firing' action of the neuron involves the generation of a spike which is implemented by an additional photonic circuit as shown in Fig. \ref{fig:neuron} (a). This circuit consists of an photonic amplifier, a circulator and a rectangular waveguide with a GST element on top initially in crystalline state. For a rectangular waveuguide with GST, the transmission is low (high) in crystalline (amorphous) state. The `read' and `write' phases for the `integration unit' and the `firing unit' alternate in successive cycles. This essentially means that during the `write' cycle of the integration unit of the neuron, a read pulse is passed through the firing unit. On the other hand, during the `read' cycle of the integration unit, the `read' pulse is passed through the ring resonators and based on the output of inteferometer and subsequent amplification, the resulting pulse attempts to write into the GST of the rectangular waveguide in the firing unit. A circulator C directs the incoming and outgoing pulses into the rectangular waveguide. When the GST elements in the integration unit are initially in crystalline state, the output of the amplifier A ($P_{amp}$) is not sufficient to amorphize the GST on rectangular waveguide and hence, a spike is not transmitted through the rectangular waveguide. However, when the membrane potential integrates, on incidence of several `write' pulse, enough to the cross the threshold, $P_{amp}$ is ensured to be high enough to amorphize the GST on the rectangular waveguide and a spike is transmitted. Once the neuron fires, a `RESET' pulse is passed to reset the states of the devices to their initial states and the membrane potential drops to the resting potential ($P_{rest}$) as shown in Fig. \ref{fig:neuron} (b). Thus, the operation of a bipolar integrate and fire neuron can be achieved using the setup described in Fig. \ref{fig:neuron}. \par 
The dynamics of the spiking neuron is primarily governed by the phase-change dynamics of GST. GST partially absorbs the wave passing through the ring waveguide and its low thermal conductivity \cite{lyeo2006thermal}causes a considerable increase in temperature. The growth of the amorphization region in the material occurs when the concerned region is above the melting temperature, which is around 877K \cite{Sebastian_2014}. For a particular incident pulse, the amorphous region heats up less than the crystalline region. Thus the change in amorphous thickness will decrease as the amorphous thickness increases. Thus, change in amorphization thickness is a function of the current state of the GST and the amplitude of the incident pulse.

% The crystallization of the GST on the ring resonator is triggered during its `write' phase by application of optical pulses. GST absorbs part of the wave passing through the ring waveguide and its low thermal conductivity \cite{lyeo2006thermal}causes a considerable increase in temperature. The growth velocity of the crystal is related to its temperature through the widely accepted model\cite{Sebastian_2014,Salinga_2013}:
% \begin{equation}
% v_g(T) = \frac{4r_{atom}k_BT}{3\pi\lambda^2R_{hyd}}.\frac{1}{\eta(T)}.\bigg[1-exp\bigg(-\frac{\Delta G(T)}{k_BT}\bigg)\bigg]
% \end{equation}
% where $r_{atom}$ is the atomic radius, $\lambda$ is the diffusional jump distance, $R_{hyd}$ is the hydrodynamic radius and $k_B$ denotes the Boltzmann constant. $\eta(T)$ is the viscosity which is estimated by and $\Delta G(T)$ is the Gibbs free energy difference between the liquid and the crystalline phase which is estimated by:
% \begin{equation}
% \Delta G(T) = \Delta H_m.\frac{T_m-T}{T_m}.\frac{2T}{T_m+T}
% \end{equation}
% where heat of fusion $\Delta H_m$ and melting temperature $T_m$ can be extracted from experimental studies \cite{Salinga_2013}. The change of amorphous thickness is a function of interfacial temperature between the crystalline and amorphous phases $T_{int}$: 
% \begin{equation}
% \frac{du_a}{dt} = -v_g(T_{int})
% \end{equation} 
\section*{Results}
\begin{table}[]
\centering
\caption{Dimensions and Material parameters}
\label{param}
\begin{tabular}{|l|l||l|l|}
\hline
\multicolumn{2}{|c||}{Dimensions}              & \multicolumn{2}{c|}{Material Parameters}                          \\ \hline
Parameters                       & Values     & Parameters                                             & Values   \\ \hline
Ring Resonator Radius ($R$)      & 6$\mu m$        & Si Refractive Index ($n_{Si}$)\cite{aspnes1983dielectric}                         & 3.5      \\
Si Waveguide Cross-section       & 0.4$\times$0.18$\mu m$ & SiO\textsubscript2 Refractive Index ($n_{SiO_2}$) \cite{malitson1965interspecimen}                    & 1.4      \\
Upper Coupling Gap ($L_{upper}$) & 0.1 $\mu m$     & c-GST Refractive Index ($n_{c-GST} +i\kappa_{c-GST}$ )\cite{kim1998variation} & 7.2+1.9i \\
Lower Coupling Gap ($L_{lower}$) & 0.1$\mu m$      & a-GST Refractive Index ($n_{a-GST}+i\kappa_{a-GST}$)\cite{kim1998variation}   & 4.6+0.18i \\
GST Length ($L_{GST}$ )          & 0.3$\mu m$      & c-GST Specific Heat,($C_{c-GST}$)                   & 217J/kg.K          \\
GST Width ($W_{GST}$ )           & 0.3$\mu m$      & a-GST Specific Heat,($C_{a-GST}$)                   & 217J/kg.K          \\
GST Thickness ($t_{GST}$ )       & 20 nm      &  c-GST Thermal Conductivity,($k_{c-GST}$)\cite{Rios_2015}  & 0.59W/m.K    \\
& & a-GST Thermal Conductivity,($k_{a-GST}$)\cite{aGST_ref_2016}&0.19
W/m.K  \\
& & c-GST Density,($\rho_{c-GST}$)\cite{Walter_2002}    & 6270 kg/m\textsuperscript{3} \\
& & a-GST Density,($\rho_{a-GST}$)\cite{Walter_2002}    & 5870 kg/m\textsuperscript{3}
								\\ \hline
\end{tabular}
\end{table}
The `write' operation of the spiking neuron is investigated using the modal profiles of the incident EM waves and the resulting temperature profiles in the GST-Si-SiO\textsubscript2 stack. The `read' operation, on the other hand, is explored from the point of view of the entire GST-ring resonator system.
The modal profile of input EM wave and subsequent heat dissipation framework was implemented in COMSOL\cite{comsol}. The temperature profiles were used to simulate the phase change characteristics of GST in MATLAB\textsuperscript\textregistered. The optical response of a ring resonator was obtained using a commercial-grade simulator Lumerical FDTD Solutions based on the finite-difference time-domain (FDTD) method\cite{lumerical}. Table. \ref{param} lists the parameters used for each simulation. 
\subsection*{Phase change dynamics of GST} 

\begin{figure*}[t]
		\centering
		\includegraphics[width=0.8\textwidth,keepaspectratio]{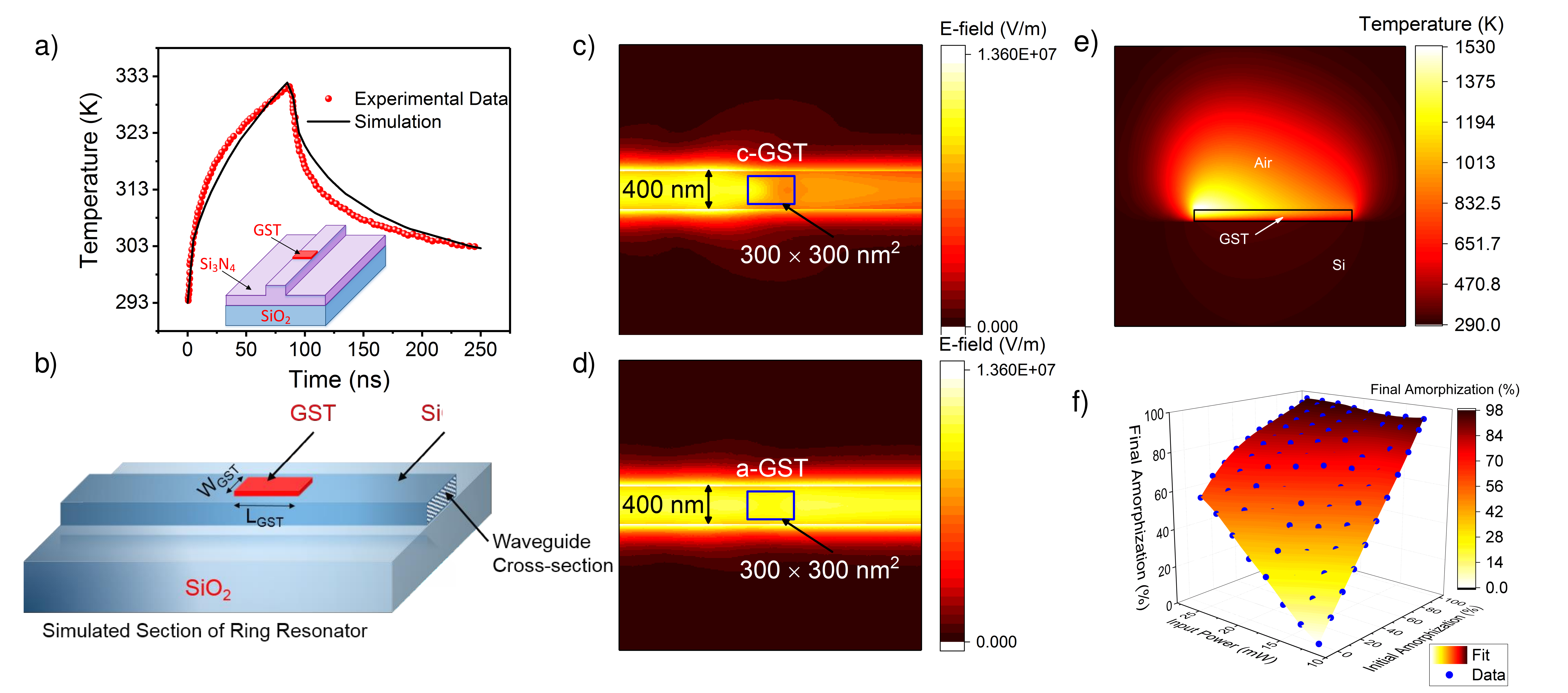}
		\vspace{-2 mm}
		\caption{(a) Experimental\cite{} benchmarking on a Si\textsubscript{3}N\textsubscript{4}-SiO\textsubscript{2} ridge-waveguide system, validating our simulation framework. (b) Simulated volume of the GST section in the ring resonator described in Fig. \ref{fig:theory} (a) delineating the different materials used. Surface electric field propagation of (c) c-GST and (d) a-GST shows significant contrast. (e) Temperature distribution along the length of cGST. (f) Plot of final percentage amorphization as a function of initial percentage amorphization and input power.}
		\vspace{-4mm}
		\label{fig:Comsol}
	\end{figure*}

The electromagnetic power absorption and subsequent temperature rise in GST is analyzed in detail using Finite Element Method (FEM) simulations in COMSOL Multiphysics. Firstly, to validate our simulation framework we simulated a GST embedded Si\textsubscript{3}N\textsubscript{4}-SiO\textsubscript{2} ridge-waveguide system and compared its transient response of temperature in GST with experimental data\cite{Rios_2015} under same excitation conditions. Fig. \ref{fig:Comsol} (a) shows good agreement between the results from our simulation and corresponding experimental data, thus validating our simulation setup.   
Next, we built a 3D model of a section of the ring resonator with GST as shown in Fig. \ref{fig:Comsol} (b) and studied the electromagnetic characteristics and subsequent temperature profiles using the validated simulation setup. The dimensions of the waveguide were fixed to ensure single fundamental mode propagation for a input optical wave of 1550 nm length. The electric field distribution at the surface of the waveguide embedded with c-GST and a-GST are shown in Fig. \ref{fig:Comsol} (c) and (d) respectively. We observe optical attenuation of - 3.71 dB in the waveguide for c-GST of 0.3 $\mu$m length and 20 nm thickness while similar dimensions of a-GST give us negligible (- 0.26 dB) attenuation. This implies strong optical absorption in c-GST and also validates the fact that it is an order of magnitude higher than that of amorphous state\cite{Rios_2015}. This property allows us to progressively amorphize our device while keeping the state of the already amorphized volume undisturbed for our chosen range of input optical power. \par 
Next, we analyze the thermal response of the GST upon optical excitation using finite element simulation. We incorporate optical heating by modeling GST as local heat source. An optical pulse of amplitude $26 mW$ and duration $200 ps$ is injected from the front facet of the waveguide. The GST is initially considered to be in crystalline state and absorbed energy in GST is taken as the heat energy for that local heat source. However, as heat is not generated uniformly within the GST volume, we designed the heat source to decrease exponentially\cite{Rios_2015} with a factor, $A = exp(-|\alpha_x|\cdot x\cdot ln(10)/10)$ along the length of the GST ($0\leq x\leq L_{GST}$) where $\alpha_x %- 12.37 dB/\mu m_{(GST)}%
$ is the optical attenuation per unit length of GST. Resulting temperature distribution at the end of the pulse is shown in Fig. \ref{fig:Comsol} (e). From inspection of this profile, an exponential temperature distribution along the GST length becomes evident. We also observe that there exists a significant portion of GST whose temperature is above the melting temperature (877K) and hence will become amorphized (e.g. $57\%$ amorphization for given conditions) after removal of optical pulse. This simulation was performed multiple times keeping the pulse width same but varying the pulse power (amplitude) and initial level of amorphization and results are plotted in Fig. \ref{fig:Comsol} (f). We find that below 12mW (200 ps) input pulse, irrespective of initial amorphization state, no further amorphization happens. Thus, we choose a input power range (26mW to 12mW) for the operation of the proposed all-photonic spiking neuron.           

\subsection*{Optical response of ring resonator}
\begin{figure*}[t]
		\centering
		\includegraphics[width=0.8\textwidth,keepaspectratio]{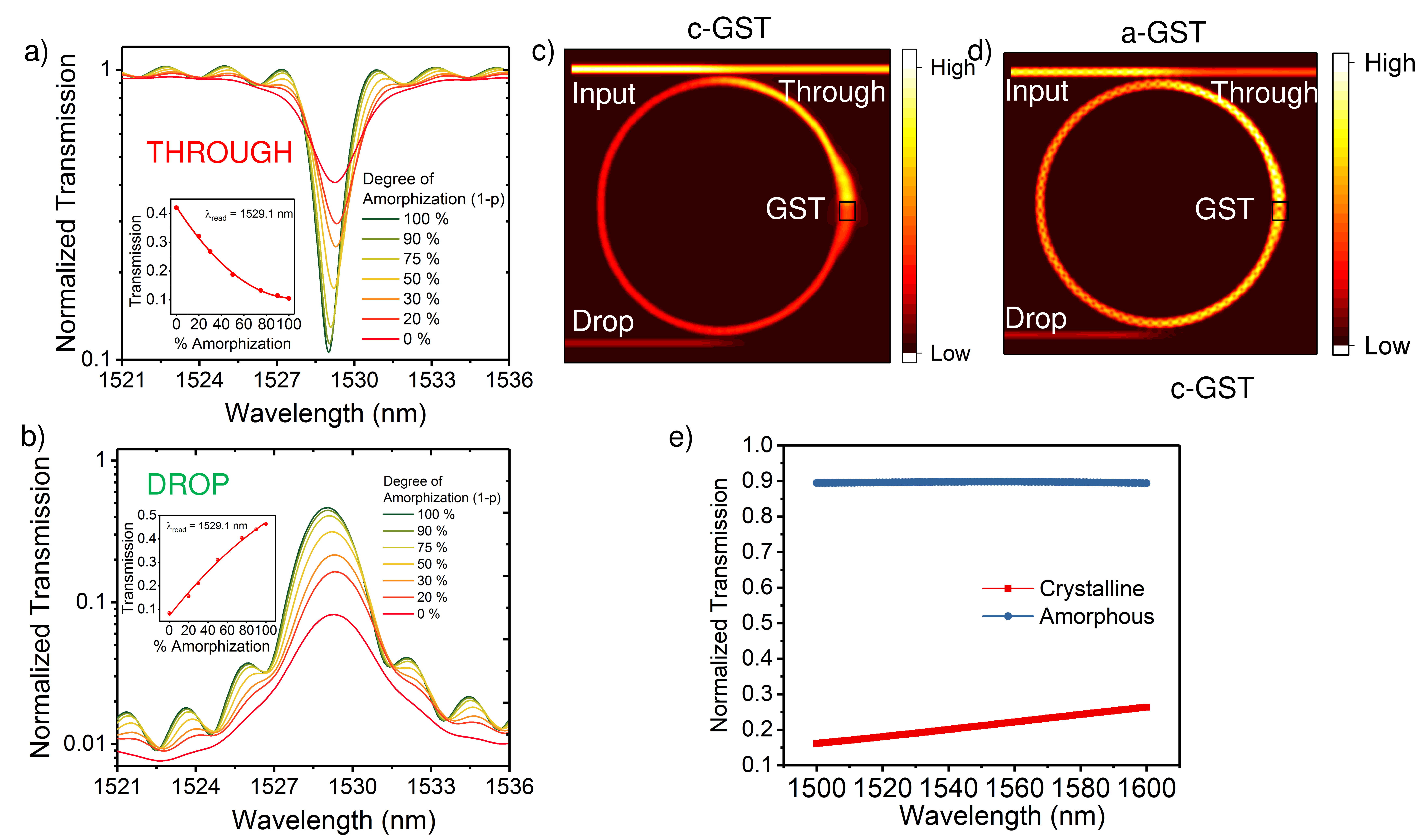}
		\vspace{-2 mm}
		\caption{Normalized Transmission at the (a) `THROUGH' and (b) `DROP' ports with increasing degree of amorphization for a particular range of frequencies including a resonance peak at $\lambda_{read} = 1529.1nm$. As the degree of amorphization increases, transmission at `THROUGH' (`DROP') port decreases (increases) thus realizing negative (positive) integration action of the neuron. (c) and (d) shows the top-view E-field distribution of a GST-embedded ring resonator for c-GST and a-GST showing higher field absorption for the former when the wave passes the GST region. (e) High contrast between c-GST and a-GST for the rectangular waveguide in the firing unit of the neuron.}
		\vspace{-4mm}
		\label{fig:fdtd}
	\end{figure*}
The `read' operation of the spiking neuron concerns with the optical response of the ring resonator or more precisely, the transmission characteristics at the `THROUGH' and `DROP' ports of the device. FDTD simulations were performed in Lumerical. Inc on a ring resonator with Si waveguides and SiO\textsubscript{2} substrate with a patch of GST on top of the ring waveguide as illustrated in Fig. \ref{fig:theory} (a). Fig \ref{fig:fdtd} (a) and (b) shows the normalized transmission at the `THROUGH'  and `DROP' ports for different amorphization levels of GST. The insets of Fig \ref{fig:fdtd} (a) and (b) show the variation of transmission at a resonant wavelength $\lambda_{read} = 1529nm$ with increasing degree of amorphization for the two ports respectively and results show consistency with our theoretical discussions above. The variation in transmission results from the decreasing absorption co-efficient ($\alpha$) as the GST amorphizes. We observe a FWHM of 1.68 (2.23) nm for a-GST and 2.97 (2.97) nm for c-GST and an extinction ratio contrast of 7.5 (6.03) dB between the fully amorphous and fully crystalline states in the `THROUGH' (`DROP') port. Fig. \ref{fig:fdtd} (c) and (d) shows the visible contrast in electric field absorption by the GST element in the ring resonator for the amorphous and crystalline states of GST for an on-resonance incident wave. The slight shift in the resonance peaks can be attributed to the minor variations in the real part of the effective refractive indices of the GST at different states, which can be expressed as\cite{Stegmaier_2016}:
\begin{equation}
\Delta \lambda_{read}  \approx \frac{\Delta n_{eff,GST}}{n_{eff,wg}}.\frac{L_{GST}}{2\pi R}
\end{equation}
These characteristics show that the outputs at the `THROUGH' and `DROP' ports decrease and increase respectively with increasing degree of amorphization which is a desirable characteristic for integration in both the positive and negative direction. We leverage this characteristic by connecting the outputs from the `THROUGH' and `DROP' ports of two devices to an interferometer, as shown in Fig. \ref{fig:neuron} (a) to obtain the resultant integration of the membrane potential as described earlier. Thus, the progressive optical responses of the ring resonator for various percentage amorphization are in agreement with the desired characteristics for the neuronal system to show integrating action. Finally, the contrast between transmission of a-GST and c-GST for a rectangular waveguide is shown in Fig. \ref{fig:fdtd} (e).

\subsection*{Spiking Neural network inferencing framework}
\begin{figure*}[t]
		\centering
		\includegraphics[width=0.8\textwidth,keepaspectratio]{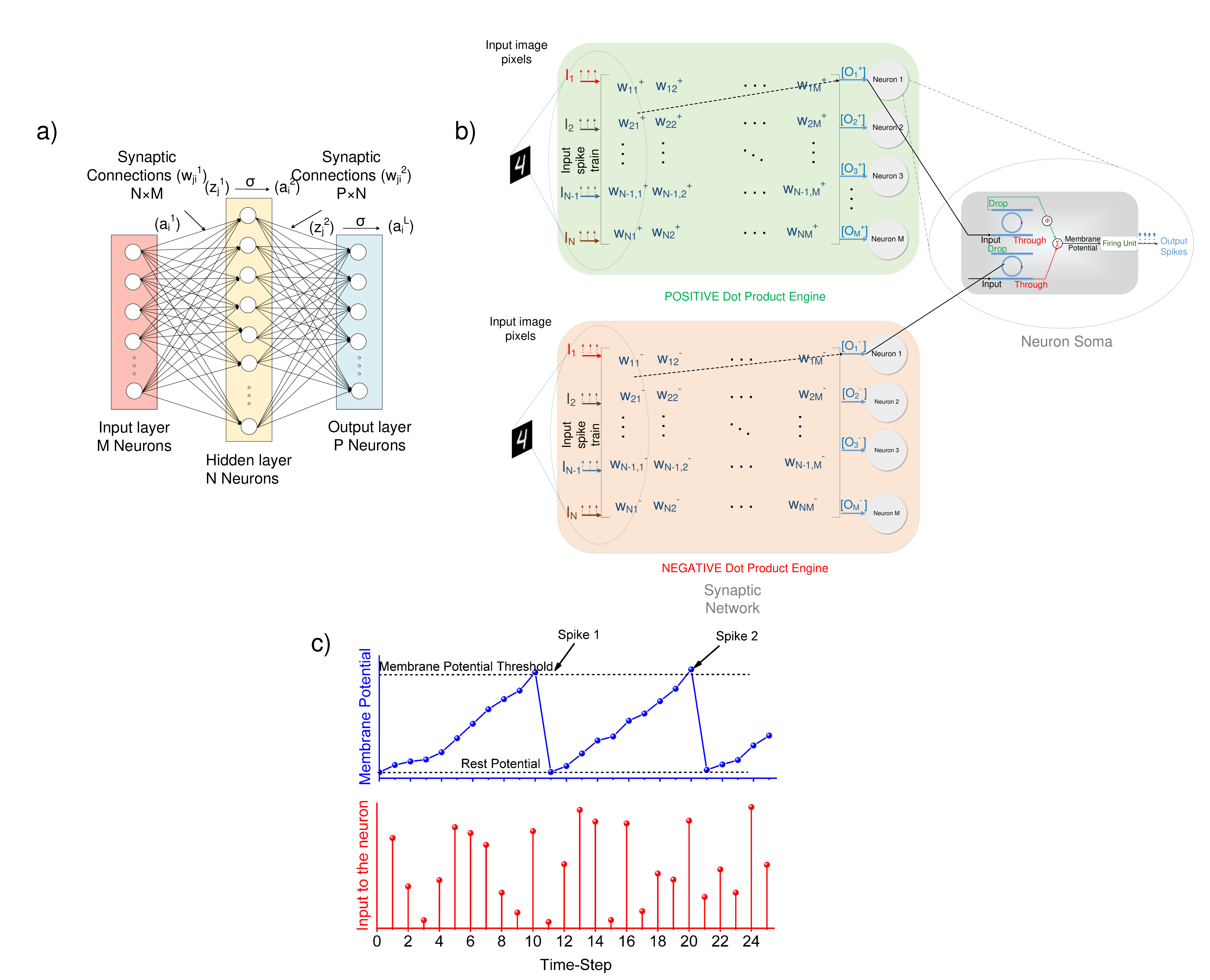}
		\vspace{-4 mm}
		\caption{(a) Fully connected ANN topology showing 3 interconnected layers, namely, the input layer, the hidden layer and the output layer\cite{chakraborty2017technology}, (b) Schematic of potential integration of an integrate-and-fire neuron in a  spiking neural network framework consisting of bipolar weights. The positive and negative weighted sums are computed using two separate dot-product engines and input to two different ring-resonators. The bidirectional integrating action of the two ports of the ring resonator is leveraged to calculate the effective membrane potential under the action of the bipolar weighted sums. Output spikes are generated when the effective membrane potential of the neuron crosses a threshold by the spike generation mechanism described. (c) The behavior of the proposed integrate-and-fire neuron in the simulated SNN showing the variation of the membrane potential under the action of incident pulses thus showing integrate and firing action. }
		\vspace{-4mm}
		\label{fig:snn}
	\end{figure*}
A neural network is comprised of multiple layers of neurons connected through synapses. The operation of any layer in a neural network involves computing the dot-product of the inputs and weights of the synapses, which gets transferred through the neuron to the next layer. To that effect, the synaptic network can be represented as a dot-product engine  that multiplies the inputs with the corresponding synaptic weights and computes a weighted sum which is received by the neuron. Such a dot-product framework can be potentially implemented by GST-based photonic synapses. Such a synapse can draw its inspiration from a GST-based on-chip photonic synapse\cite{cheng2017chip} recently proposed. The proposed integrate-and-fire spiking neuron can be integrated with these photonic synapses in an all-photonic implementation of a spiking neural network. To analyze the performance of such an all-photonic neural network, we built a device to algorithm framework by mapping the device characteristics to implement the proposed neuron in an algorithm level neural network inferencing setup. Such a system-level simulation is quintessential to validate the operation of the proposed integrate-and-fire neuron. For the current analysis, we assume ideal operation of the dot-product engine. We consider a fully connected network consisting of 3 layers, the input layer, the hidden layer and the output layer as shown in Fig. \ref{fig:snn} (a). In such a network, each neuron receives inputs from all the neurons of the previous layer. We study the performance of the aforementioned fully connected neural network in a standard handwritten digit recognition task based on the MNIST dataset\cite{lecun-mnisthandwrittendigit-2010}. The MNIST dataset consists  of 60000 training images and 10000 testing images.  The weights of the synapses are trained using the Backpropagation algorithm\cite{hecht1992theory} as in case of traditional Artificial Neural networks (ANN). During inferencing, we use a conversion mechanism \cite{diehl2015fast} from ANN to SNN where the neurons with `ReLU'\cite{nair2010rectified} activation functions in the ANN are replaced by the proposed integrate-and-fire neurons. The dependence of final state of the device on the input and initial state of the device as shown in Fig. \ref{fig:Comsol} (f) was used to determine the state of each neuron after each time-step. Then, the transmission characteristics of the ports of the ring resonators in the proposed neuron as shown in Fig. \ref{fig:fdtd} (a) and (b) was used to determine the final membrane potential of each neuron. Each pixel of a 28$\times$28 input image is divided into a stream of spikes whose frequency is proportional to the pixel intensity. The proposed integrate-and-fire neurons receive the dot product of the input spikes in a certain time-step $t$ and the corresponding weights of synapses connecting the neuron and the inputs as shown in Fig. \ref{fig:snn} (b). Upon receiving the dot product stimulus, the neurons integrate its membrane potential at that time-step. Mathematically, for $j^{th}$ neuron, this can be represented similar to Eqn.6:
\begin{equation}
V_j[t] = V_j[t-1] + \sum_i I_i[t]w_{ij}
\end{equation}
where $V_j[t]$ is the internal state or the membrane potential of the $j^{th}$ neuron at time $t$, $I_i[t]$ is the $i^{th}$ input at time $t$, $w_{ij}$ is the weight of the synapse connecting the $i^{th}$ input to the $j^{th}$ neuron. The details of the synaptic network implementation in the photonic domian will be a future course of study, however, similar concepts have been well-explored in the electrical domain \cite{sengupta2017encoding}. Any synaptic network is essentially a dot-product engine performing element-wise multiplication of the inputs and the synaptic weights. Such a dot-product engine receives an N-dimensional input vector and provides an M-dimensional output vector which can be mathematically represented as:
\begin{equation}
\begin{bmatrix}O_1\\O_2\\\vdots\\O_M\end{bmatrix} = \begin{bmatrix}I_1&I_2&\dots &I_N\end{bmatrix}\begin{bmatrix}w_{11} & w_{12} & \dots & w_{1M}\\w_{21} & w_{22}& \dots & w_{2M}\\\vdots & \vdots &\ddots &\vdots\\w_{N1} & w_{N2} & \dots & w_{NM}\end{bmatrix} 
\end{equation}
where $[w_{ij}]$ is a $N\times M$ weight matrix.\par
To account for weights of either polarity, we represent the weights in two different dot-product engines as shown in Fig. \ref{fig:snn} (b). We can interpret the weight $w_{ij}$ to possess a positive and negative component:
\begin{gather}
w_{ij} = w_{ij}^+-w_{ij}^-\\
w_{ij}^- = |w_{ij}|, w_{ij}^+ = 0, \textnormal{when }w_{ij}<0\\
w_{ij}^+ = w_{ij}, w_{ij}^- = 0, \textnormal{when } w_{ij}>0
\end{gather}
This gives us two matrices $W^+ = [w_{ij}^+]$ and $W^- = [w_{ij}^-]$. These matrices are represented in the dot-product engines such that they return the corresponding dot products:
\begin{gather}
O_j^{+} = \sum_i{I_iw_{ij}^{+}}\\
O_j^{-} = \sum_i{I_iw_{ij}^{-}}
\end{gather}
The positive and negative integrating ring resonators in the proposed neuron take these inputs separately and integrate in opposite direction such that the resulting integration mimics the desired integration that a biological neuron performs, given by Eqn. 7 because $\sum_i I_iw_{ij} =  \sum_i{I_iw_{ij}^+}-\sum_i{I_iw_{ij}^-}$. The resulting membrane potential is fed to a Firing Unit as described in Fig. \ref{fig:neuron} (a). A behavorial model of the SNN inferencing framework described above was simulated using the MATLAB Deep Learning Toolbox \cite{palm2012prediction} using a well-explored network topology \cite{diehl2015fast}. Fig. \ref{fig:snn} (c) shows the progression of the membrane potential of the proposed integrate-and-fire neuron in the hidden layer of the simulated SNN under the action of weighted incident spikes with time. The magnitude of the weighted incident spikes is essentially equal to $\sum_i I_i[t]w_{ij}$ for the $j^{th}$ neuron at time-step $t$. It can be observed that once the membrane potential of the neuron reaches its threshold, it goes back to its rest potential. In the process, it generates a spike that gets fed to the next layer. The same integration process happens in case of the output layer neurons as well and the spike activities of all the neurons are monitored. The 10 output layer neurons correspond to the 10 classes of image being classified. The neuron with the highest spiking activity over a number of time-steps is compared with the test image label and if it matches with the neuron number, the image is classified correctly. This device to system level analysis helps us validate the operation of the proposed integrate-and-fire neuron. The accuracy of recognition was calculated to be 98.06\% after 25 time-steps on the testing set. The accuracy suffers a 0.24\% degradation with respect to the testing accuracy (98.3\%) of a SNN based on an ideal integrate-and-fire neurons. This can be attributed to the non-linear transmission characteristics shown in Fig. \ref{fig:fdtd} and the dependence of the final state on the initial state of the device. Such device inaccuracies can be accounted for by modifying the training algorithm \cite{chakraborty2017technology}.

\par
The important metrics for performance evaluation on a neuromorphic hardware system are energy efficiency and speed. To that effect, the energy and delay performance of the proposed neuron merits discussion. Each `write' cycle is considered to be $1.5 ns$ and each `read' cycle for the proposed neuron was considered to be $500 ps$. The durations of the `read' and `write' pulses were $200 ps$. The additional times in the `write' and `read' cycles is to ensure that the GST temperature settles to its initial value after the excitation. The `write' times are constrained by the transient response of GST to an amorphization pulse, which is shown to achieve times as low as $200 ps$, experimentally \cite{Stegmaier_2016} when excited with $1 ps$ pulses. The average energy of a `write' step considered for the simulation of the neural network was 4 pJ per neuron per time-step whereas the average `read' energy was 1 pJ per neuron per time-step. The energy consumption in the `write' cycles of the neuron can be further reduced by optimizing the feature size of the GST element. PCM devices of similar feature sizes\cite{lee2009architecting,wong2010phase} in the electrical domain can consume upto 14-19 pJ of `write' energy while operating at speeds of 40-100ns. Writing into the GST through evanscent coupling with photonic waveguides thus achieves a higher energy efficiency and speed, thus promising to rekindle the viability of PCMs for fast neuromorphic processing. 

\section*{Discussion}

Neuromorphic engineering has evolved heavily from its dawn as researchers have explored various kinds of technologies to mimic the functionality of the brain on an energy-efficient hardware platform. In the electrical domain, such technologies have been demonstrated to possess limitations such as speed, energy, process integration etc. Phase change materials, in particular, have hit the scaling bottleneck where further improvements in energy-efficiency would require reducing `write' speeds significantly. To beat CMOS in terms of energy-efficiency a 10$\times$ reduction\cite{rajendran2013specifications} in current pulse amplitude or increase in pulse duration is necessary. As a solution, we propose an all-photonic integrate-and-fire neuron based on the phase change dynamics of GST which promises to achieve `write' speeds of sub-ns orders. To the best of our knowledge, this is the first demonstration of a biologically plausible spiking neuron in the photonic domain involving phase change materials. We also showed that the proposed neuron can be potentially integrated with synapses in an all-photonic spiking neural network inferencing framework without any significant drop in classification performance. The proposed design opens up a host of possibilities for future implementations of all-photonic SNNs. By modulating the resonant wavelength by varying dimensions offers us the opportunity of wavelength multiplexing in an all-photonic spiking neural network. This offers substantial benefits such as elimination of cross-talk between neighboring neural elements thus allowing the provision of a denser network and in addition, could possibly allow us to implement larger networks on the same chip. With the recent advances in Photonic Neuromorphic, the proposed integrate-and-fire neuron fills the void of an all-photonic neuron that can be interfaced with photonic synapses\cite{cheng2017chip} to build a truly integrated all-photonic neuromorphic system that leverages the aforementioned advantages of photonic devices to perform ultrafast neuromorphic computation. \par 
% \section*{Methods}

% Topical subheadings are allowed. Authors must ensure that their Methods section includes adequate experimental and characterization data necessary for others in the field to reproduce their work.

\bibliography{sample}

% \noindent LaTeX formats citations and references automatically using the bibliography records in your .bib file, which you can edit via the project menu. Use the cite command for an inline citation, e.g.  \cite{Figueredo:2009dg}.

\section*{Acknowledgements}

The work was supported in part by, ONR-MURI program, the National Science Foundation, Intel Corporation and by the DoD Vannevar Bush Fellowship.

\section*{Author contributions statement}

I.C. and K.R. conceived the study. I.C. conceived the necessary simulations,  I.C. and G.S. conducted the simulations, I.C., G.S and A.S analyzed the results. All authors reviewed the manuscript. 

\section*{Additional information}
 
The authors declare no competing interests.

\end{document}